\documentclass[aps,prl,twocolumn,superscriptaddress]{revtex4}

\usepackage[dvips]{graphicx}
\usepackage{times,bm} 
\usepackage{amsmath}

\newcommand{\bG}{\mbox{\boldmath $G$}}

\newcommand{\bq}{\mbox{\boldmath $q$}}
\newcommand{\br}{\mbox{\boldmath $r$}}

\begin{document}

\title{{\em Ab initio} Derivation of Low-Energy Model for $\kappa$-ET Type Organic Conductors}

\author{Kazuma Nakamura} 
\thanks{Electronic mail: kazuma@solis.t.u-tokyo.ac.jp}
\affiliation{Department of Applied Physics, University of Tokyo, 7-3-1 Hongo Bunkyo-ku, Tokyo 133-8656, Japan} 
\affiliation{JST, CREST, 7-3-1 Hongo, Bunkyo-ku, Tokyo 113-8656, Japan}

\author{Yoshihide Yoshimoto} 
\affiliation{Institute for Solid State Physics, 
 University of Tokyo, 5-1-5 Kashiwanoha, Kashiwa, Chiba 277-8531, Japan} 

\author{Taichi Kosugi}
\affiliation{Department of Physics, 
 University of Tokyo, 7-3-1 Hongo, Bunkyo-ku, Tokyo 113-0033, Japan} 

\author{Ryotaro Arita} 
\affiliation{Department of Applied Physics, University of Tokyo, 7-3-1 Hongo Bunkyo-ku, Tokyo 133-8656, Japan}
\affiliation{JST, CREST, 7-3-1 Hongo, Bunkyo-ku, Tokyo 113-8656, Japan}

\author{Masatoshi Imada} 
\affiliation{Department of Applied Physics, University of Tokyo, 7-3-1 Hongo Bunkyo-ku, Tokyo 133-8656, Japan}
\affiliation{JST, CREST, 7-3-1 Hongo, Bunkyo-ku, Tokyo 113-8656, Japan}

\date{\today}

\begin{abstract}

We derive effective Hubbard-type Hamiltonians of $\kappa$-(ET)$_2X$, using an {\em ab initio} downfolding technique, for the first time for organic conductors. They contain dispersions of the highest occupied Wannier-type molecular orbitals with the nearest neighbor transfer $t$$\sim$0.067 eV for a metal $X$=Cu(NCS)$_2$ and 0.055 eV for a Mott insulator $X$=Cu$_2$(CN)$_3$, as well as screened Coulomb interactions. It shows unexpected differences from the conventional extended H\"uckel results, especially much stronger onsite interaction $U$$\sim$0.8 eV ($U/t$$\sim$12-15) than the H\"uckel estimates ($U/t$$\sim$7-8) as well as an appreciable longer-ranged interaction. Reexamination on physics of this family of materials is required from this realistic basis.

\end{abstract}


\maketitle
 
Organic conductors with ET molecules, (ET)$_2X$ with a number of choices of anions $X$, offer a variety of prototypical behaviors of strongly correlated electron systems with two-dimensional (2D) anisotropies \cite{Kanoda}. Examples range from correlated metals with superconductivity at low temperatures to Mott insulators either with a quantum spin liquid or with antiferromagnetic, charge-density or spin-Peierls orders.  Intriguing Mott transitions are also found. 
They are all in front of recent active research for unconventional quantum phases and quantum critical phenomena in nature, while their essences of physics are still under strong debates.  

In particular, an unconventional nonmagnetic Mott-insulating phase is found near the Mott transition in the $\kappa$-type structure of ET molecules, $X$=Cu$_2$(CN)$_3$ referred to as $\kappa$-CN, where no magnetic order is identified down to the temperature $T$=0.03 K, four orders of magnitude lower than the antiferromagnetic spin-exchange interaction $J$$\sim$250 K \cite{Shimizu}. The emergence of the quantum spin liquid near the Mott transition has been predicted in earlier numerical studies \cite{Kashima,Morita,Mizusaki}, while the full understanding of the spin liquid needs more thorough studies. It is also crucially important to elucidate the real relevance of the theoretical findings to the real $\kappa$-ET compounds.  Most of numerical \cite{Kyung} and theoretical \cite{Lee} studies have also been performed for a simplified single-band 2D Hubbard model based on an empirical estimate of parameters combined with extended H\"uckel calculations \cite{Mori,Saito}.  A more realistic description of $\kappa$-ET compounds is certainly needed beyond the empirical model.  

Another fundamental finding achieved in this series of compound is the unconventional Mott transition found for $X$=Cu[N(CN)$_2$]Cl under pressure \cite{Kagawa}.   
The novel universality class of the Mott transition is in good agreement with the marginal quantum criticality at the meeting point of the symmetry breaking and topological change \cite{Imada1,Imada2,Misawa1,Misawa2}.
 Because of its significance to the basic understanding on the physics of quantum criticality, the relevance of theoretical concept to the experimental observation needs to be further examined on the realistic and first-principles grounds.  Furthermore, an unconventional superconductivity is found in the metallic sides of these compounds at low temperatures ($T$$<$$T_c$$\sim$10-13K), where the mechanism is not clearly understood yet \cite{Geiser,Cu2CN3}.  
These outstanding properties of $\kappa$-ET compounds encourage systematic studies based on realistic basis. As mentioned above, however, the first-principle studies are limited \cite{DFT_ET} and most of the studies so far were performed using the empirical models inferred from the H\"uckel studies.   

The purpose of this letter is to derive {\em ab initio} effective Hamiltonian of the real $\kappa$-ET compounds to serve in establishing a firm basis for studies on the unconventional phenomena and open issues. 
The present work is the first challenge aiming at {\em ab initio} model construction of organic conductors containing a large number of atoms with four complex ET molecules in a unit cell. 
 We derive models of two contrasting compounds, spin-liquid $\kappa$-CN and superconducting compound $X$=Cu(NCS)$_2$ abbreviated as $\kappa$-NCS \cite{CuSCN}, to get insights into the whole series of $\kappa$-(ET)$_2X$ compounds from metals to Mott insulators. 
Our {\em ab initio} results indicate a substantial difference from the previous simple extended H\"uckel results \cite{Shimizu,Saito,Mori};  
in particular, a large ratio of the onsite Coulomb repulsion to the nearest neighbor transfer $U/t$$\sim$12 even in metallic $\kappa$-NCS compared to $\sim$7 in the H\"uckel estimate requires reexamination of the model studies. 
  
A reliable first-principle framework has been proposed for the methods to derive effective low-energy Hamiltonians of real materials. 
This treatment consists of {\em ab initio} density-functional calculations of the global electronic band structure and a subsequent downfolding procedure by elimination of degrees of freedom far away from the Fermi level \cite{Aryasetiawan,Solovyev,Imai,Otsuka}.  
Recently, the technique was combined with a framework of maximally localized Wannier orbital (MLWO) \cite{Miyake,Nakamura}.
The MLWO \cite{Ref_MLWF} has a distinct computational advantage, because it enables construction of localized orbitals even for the present complex molecular solid \cite{SOD}.
Below we will demonstrate how the Wannier-framework downfolding and the derivation of effective models have successfully been worked out.    

Now, we consider {\em ab initio} derivations of a single-band extended Hubbard Hamiltonian describing the low-energy electronic property of $\kappa$-(ET)$_2$X.
For the system, the basis of the Hamiltonian is the Wannier functions associated with antibonding states of the highest occupied molecular orbitals (HOMOs) of ET molecules forming a dimer. 
The explicit form of this Hamiltonian is given by 
\begin{eqnarray}
\mathcal{H} 
= \sum_{\sigma} \sum_{ij} t_{ij} a_{i \sigma}^{\dagger} a_{j \sigma}            +\frac{1}{2} \sum_{\sigma \rho} \sum_{ij}  V_{ij} a_{i \sigma}^{\dagger} 
 a_{j \rho}^{\dagger} a_{j \rho} a_{i \sigma}, \label{H_Hub}
\end{eqnarray} 
where $a_{i \sigma}^{\dagger}$ ($a_{i \sigma}$) is a creation (annihilation) operator of an electron with spin $\sigma$ in the Wannier orbital localized at the $i$th ET dimer. 
The $t_{ij}$\ parameters are written by 
\begin{eqnarray}
t_{ij} = \langle \phi_{i} | \mathcal{H}_0 |  \phi_{j} \rangle \label{t_ij} 
\end{eqnarray}
with $| \phi_{i} \rangle =a_{i}^{\dagger}|0\rangle$
and $\mathcal{H}_0$ being the one-body part of $\mathcal{H}$.
The $V_{ij}$ values are screened Coulomb integrals in the Wannier orbital, expressed as 
\begin{eqnarray}
V_{ij}= \int \int d\br d\br' \phi_{i}^{*}(\br) \phi_{i}(\br) 
W(\br,\br') \phi_{j}^{*}(\br') \phi_{j}(\br'), \label{V_ij}
\end{eqnarray} 
where $W(\br,\br')$ is a screened Coulomb interaction. 
The $V_{ij}$ at $i$ = $j$ corresponds to onsite Hubbard repulsion $U$. 
Here, $W$ is practically calculated in the reciprocal space as 
\begin{eqnarray}
 W(\br,\br')=\frac{4 \pi}{\Omega} \sum_{{\bf q} {\bf G} {\bf G}'} 
 \frac{e^{i ({\bf q}+{\bf G}) {\bf r}}}{| \bq+\bG |} 
 \epsilon_{{\bf G} {\bf G}'}^{-1}(\bq) 
 \frac{e^{-i ({\bf q}+{\bf G}') {\bf r'}}}{| \bq+\bG'|}, 
  \label{eq:W}
\end{eqnarray} 
where $\bq$ is a wave vector in the first Brillouin zone and $\bG$ is a reciprocal lattice vector.
The crystal volume is $\Omega$ and 
$\epsilon_{{\bf G}{\bf G}'}^{-1}(\bq)$ is the
inverse dielectric matrix which
is related to the irreducible polarizability $\chi$ by 
$\epsilon_{{\bf G}{\bf G}'}(\bq) 
= \delta_{{\bf G}{\bf G}'} - v(\bq+\bG) \chi_{{\bf G}{\bf G}'}(\bq)$,  
where $v(\bq)$=4$\pi/\Omega |\bq|^2$ is the 
bare Coulomb interaction. 
The polarization matrix $\chi_{{\bf G} {\bf G}'}(\bq)$ with a random phase approximation (RPA) is calculated as
\begin{eqnarray}
\chi_{{\bf G} {\bf G}'}(\bq)&=&
{\sum_{{\bf k}} \sum_{\alpha \beta}}
\langle \psi_{\alpha {\bf k}+{\bf q}} 
| e^{i ({\bf q}+{\bf G}) {\bf r}} 
| \psi_{\beta {\bf k}} \rangle  \nonumber \\
&& \times
\langle \psi_{\beta {\bf k}} 
| e^{-i ({\bf q}+{\bf G}) {\bf r}} | 
\psi_{\alpha {\bf k}+{\bf q}} \rangle 
\frac{ f_{\alpha {\bf k}+{\bf q}} - f_{\beta {\bf k}}} 
{E_{\alpha {\bf k}+{\bf q}} - E_{\beta {\bf k}}}, \label{eq:chi} 
\end{eqnarray}
where $\psi_{\alpha {\bf k}}$, $E_{\alpha {\bf k}}$, and $f_{\alpha {\bf k}}$ are the Bloch state, its energy, and the occupancy, respectively. 
The calculation of $\chi$ is straightforward with Eq.~(\ref{eq:chi}) but there is a notice when we consider the model construction. 
The screened interaction put in the Hubbard model [$V_{ij}$ in Eq.~(\ref{H_Hub})] should not include screening formed in a target band of the model \cite{Aryasetiawan,Solovyev}. 
This screening should be considered at the step of solving the effective model and, at the downfolding stage, we must exclude the target-band screening effects in the calculation of $V_{ij}$.  
In RPA, this exclusion is easily implemented, since the RPA polarization in Eq.~(\ref{eq:chi}) is written in terms of the sum of the band pairs $\alpha\beta$ associated with individual transitions.
We thus generate the polarization function with cutting the transitions between the target bands and then evaluate the screened interaction $W$ by using this polarization function.  
Finally, we compute the $V_{ij}$ parameters as the Wannier matrix elements of the $W$ interaction. 
The practical details for the matrix evaluation can be found in Ref.~\onlinecite{Nakamura}. 

Our {\em ab initio} calculations were performed with {\em Tokyo Ab initio Program Package} \cite{Ref_TAPP}.  
With this program, density-functional calculations \cite{Ref_DFT} 
with the generalized-gradient-approximation (GGA) 
exchange-correlation functional \cite{Ref_PBE96}
were performed using a plane-wave basis set and the 
Troullier-Martins norm-conserving pseudopotentials \cite{Ref_PP1} 
in the Kleinman-Bylander representation \cite{Ref_PP2}.
The energy cutoff in the band calculation was set to 36 Ry and a 3$\times$5$\times$5 $k$-point sampling was employed. 
The experimental crystal-structure data were taken from Ref.~\onlinecite{CuSCN} for $\kappa$-NCS and Ref.~\onlinecite{Cu2CN3} for $\kappa$-CN.  
The polarization function was expanded in plane waves with an energy cutoff of 3.5 Ry and the total number of bands considered in the polarization calculation was set to 634 for $\kappa$-NCS and 612 for $\kappa$-CN. 
[The total number of the occupied (unoccupied) states is 234 (400) for $\kappa$-NCS and 242 (370) for $\kappa$-CN.]
The Brillouin-zone integral over wavevector was evaluated by the tetrahedron method. 
The additional terms in the long-wavelength polarization function due to nonlocal terms in the pseudopotentials was explicitly considered following Ref.~\onlinecite{Louie}. 
The singularity of the Coulomb interaction in the $\bq$$\to$0 limit, in the evaluation of the elements $V_{ij}$, was carefully handled on the manner described in Ref.~\onlinecite{Louie}.

\begin{figure}[h]
\includegraphics[width=0.45\textwidth]{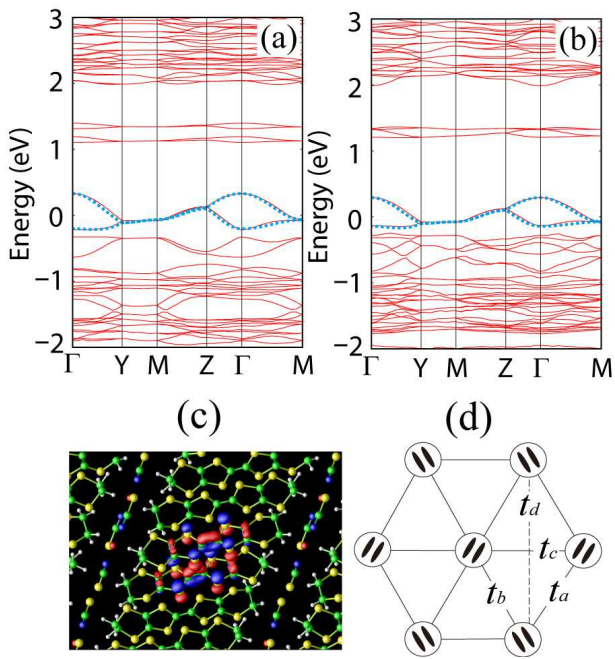} 
\caption{Calculated {\em ab initio} GGA band structures (red line) of  $\kappa$-(BEDT-TTF)$_2$Cu(NCS)$_2$ (a) and $\kappa$-(BEDT-TTF)$_2$Cu$_2$(CN)$_3$ (b). The zero of energy is the Fermi level. The blue dotted dispersions are obtained by the four transfer parameters listed in TABLE I. (c) Calculated maximally localized Wannier functions of $\kappa$-(BEDT-TTF)$_2$Cu(NCS)$_2$. The amplitudes of the contour surface are +1.5/$\sqrt{v}$ (blue) and $-$1.5/$\sqrt{v}$ (red), where $v$ is the volume of the primitive cell. S, C, H, N, and Cu nuclei are illustrated by green, yellow, silver, blue, and red spheres, respectively. (d) Schematic diagram for transfer network in the triangular lattice.}
\label{fig:band}
\end{figure}

We show in Fig.~\ref{fig:band} {\em ab initio} GGA band structures of $\kappa$-NCS (a) and $\kappa$-CN (b). 
The {\em ab initio} results clearly show that the band near the Fermi level (energy zero) is totally isolated from other bands, thus justifying employing this band as the target band of the extended Hubbard model.
The entangled bands below $-$0.5 eV and above $+$1 eV are associated with deeper and upper states of the ET molecules, as well as the anion electronic structures. 
We find that the bandwidth of $\kappa$-NCS estimated as 0.56 eV is larger than that of $\kappa$-CN (0.45 eV), which generates a discernible difference in transfer integrals of the two compounds (see below). 
We also note that the {\em ab initio} dispersion of the target band is basically similar to the result of the extended H\"uckel model but our bandwidth is somewhat smaller than the H\"uckel results of 0.57 eV for $\kappa$-NCS and 0.50 eV for $\kappa$-CN \cite{Saito}. 

\begin{table}[h] 
\caption{List of the parameters in the single-band extended Hubbard Hamiltonian in Eq.~(\ref{H_Hub}) for $\kappa$-(ET)$_2X$.} 
\centering 
\begin{tabular}{lr@{\ \ \ \ \ \ }r} \hline \hline
 & \multicolumn{1}{c}{$X$=Cu(NCS)$_2$} & \multicolumn{1}{c}{$X$=Cu$_2$(CN)$_3$} \\ \hline
   $t_a$ (meV)  & $-$64.8   & $-$54.5  \\ 
   $t_b$ (meV)  & $-$69.3   & $-$54.7  \\  
   $t_c$ (meV)  &    44.1   &    44.1  \\ 
   $t_d$ (meV)  & $-$11.5   & $-$ 6.8  \\ 
   $U  $ (eV)   &     0.83   &    0.85 \\ 
   $\lambda$ (eV$^{-1}$$\AA^{-1}$)   &     5.1    &    5.0  \\ \hline \hline
\end{tabular} 
\label{PARAM} 
\end{table}

Figure \ref{fig:band}~(c) visualizes our calculated maximally localized Wannier orbitals for $\kappa$-NCS, for the isolated band discussed above. 
An initial guess for generating the Wannier function is the $p$-type Gaussian sitting on the dimer center. 
The resulting plot for the Wannier orbital clearly exhibits that the basis state of the low-energy effective model of the $\kappa$-NCS compound is the antibonding states consisting of a linear combination of the ET HOMOs in the dimer \cite{Kuroki}.
We note that there is no discernible difference between the MLWO of $\kappa$-NCS and that of $\kappa$-CN. 

The transfer integrals in Eq.~(\ref{t_ij}) calculated as matrix elements of the Kohn-Sham Hamiltonian in the Wannier orbital are listed in TABLE I.
We list only the four parameters ($t_a$, $t_b$, $t_c$, and $t_d$) and the absolute value of other transfers are all less than 5 meV.
The schematic diagram for the transfer network is illustrated in Fig.~\ref{fig:band}~(d). 
The band dispersion calculated with the four transfers is given as blue dots in Figs.~\ref{fig:band} (a) and (b) and, from the figures, we see that the original band structure is well reproduced with these four transfers. 
Considering $t_a$$\sim$$t_b$$\equiv$$t$ (nearest-neighbor transfer), $t_c$$\equiv$$t'$ (next-neighbor transfer), and $t_d$$\sim$0, a rough estimate for geometrical frustration $|t'/t|$ is 0.66 for $\kappa$-NCS and 0.80 for $\kappa$-CN, thus indicating that the degree of frustration in $\kappa$-CN is larger than that in $\kappa$-NCS, while the estimates are still far from the perfect situation $|t'/t|$=1 inferred by the extended H\"uckel result for $\kappa$-CN \cite{Saito}.  

\begin{figure}[h]
\includegraphics[width=0.45\textwidth]{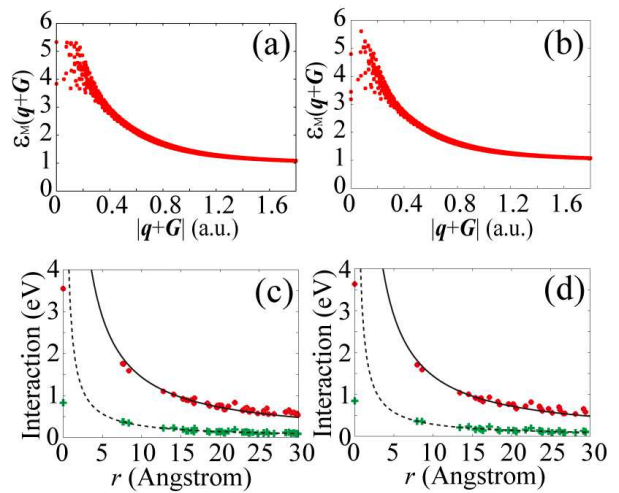} 
\caption{Calculated macroscopic dielectric functions of $\kappa$-(BEDT-TTF)$_2$Cu(NCS)$_2$ (a) and $\kappa$-(BEDT-TTF)$_2$Cu$_2$(CN)$_3$ (b) as a function of $|\bq+\bG|$. Calculated screened Coulomb interactions of $\kappa$-(BEDT-TTF)$_2$Cu(NCS)$_2$ (c) and $\kappa$-(BEDT-TTF)$_2$Cu$_2$(CN)$_3$ (d) as a function of the distance between the centers of maximally localized Wannier orbitals. The red and green dots represent the bare and screened interactions, respectively. The solid and dotted curves denotes $1/r$ and $1/(\lambda r)$, where a decay constant $\lambda$ was determined by the fitting to the {\em ab initio} data (see TABLE~\ref{PARAM}).}
\label{fig:epsilon}
\end{figure}

We next show in Fig.~\ref{fig:epsilon} calculated {\em ab initio} macroscopic dielectric function for $\kappa$-NCS (a) and $\kappa$-CN (b).
The plots for $\epsilon_{\rm M}(\bq+\bG)$$\equiv$$1/\epsilon_{{\bf G}{\bf G}}^{-1}(\bq)$ with $\epsilon_{{\bf G}{\bf G}}(\bq)$ being the dielectric functions calculated with constrained RPA are illustrated as a function of $|\bq+\bG|$.   
We see that, in the limit $\bq+\bG$$\to$0, $\epsilon_{\rm M}(\bq+\bG)$ converges to the two finite values. 
It should be noted here that the metallic screening effects are excluded in the constrained RPA calculations.
 In $\kappa$-NCS, the higher value of 5.3 is the constrained-RPA dielectric constant for the electric field perpendicular to the two-dimensional layer, while the smaller value of 3.8 is the in-plane value.
 This is characteristic feature of the anisotropic system.
For $\kappa$-CN, the higher dielectric constant is 4.8 and the smaller one is around 3.4. 

The data quality of the dielectric function critically affects the accuracy of the screened Coulomb interaction of $W$ [Eq.~(\ref{eq:W})]. 
In general, the dielectric function severely depends on the number of the conduction bands considered in the calculations \cite{Louie}. 
In the ET compounds, the conduction states were found to be fully continuous and highly-entangled.
So, the convergence of the dielectric function of this system was checked by increasing the number of the conduction bands. 
We found that the convergence requires more than 350 conduction bands, which corresponds to considering the excitation up to 16 eV above the Fermi level in the polarization calculation. 

The lower two panels of Fig.~\ref{fig:epsilon} plots our calculated {\em ab initio} screened Coulomb interaction $V_{ij}$ [Eq.~(\ref{V_ij})] with constrained RPA, denoted by green dots, as a function of the distance between the centers of the MLWOs; $r$=$|\langle \phi_{i}| \br | \phi_{i} \rangle$$-$$\langle \phi_{j}| \br | \phi_{j} \rangle |$. 
The panels (c) and (d) show the results for $\kappa$-NCS and $\kappa$-CN, respectively. 
The $V_{ij}$ decays as an isotropic function of $1/(\lambda r)$ (dotted line) beyond the nearest-neighbor distance ($\ge$7 $\AA$), where $\lambda$ is a decay constant determined by the fitting procedure to the {\em ab initio} data.
 We found that the resulting $\lambda$ is 5.1 for $\kappa$-NCS and 5.0 for $\kappa$-CN. 
For comparison, we also plot bare Coulomb interactions as red dots, which should decay as $1/r$ (solid line) \cite{Ducasse}. 
We see that the screening by the constrained RPA reduces the bare onsite Coulomb repulsions to the values less than a quarter.
The onsite Hubbard $U$ thus obtained is found to be 0.83 eV for $\kappa$-NCS and 0.85 eV for $\kappa$-CN. 

Our derived low-energy models differ from the conventional ones considered so far for the $\kappa$-type compounds. 
The {\em ab initio} calculations give a larger ratio $U/t$$\sim$12-15 than the H\"uckel results $\sim$7-8 \cite{Shimizu,Saito,Mori}.
Also, the ratio of the nearest neighbor repulsion $V$ to the onsite Hubbard $U$  is 0.45 for $\kappa$-NCS and 0.43 for $\kappa$-CN, which are definitely larger than the cases of transition-metal oxides [for example, SrVO$_3$ ($V/U$$\sim$0.2)], an iron-based oxypnictide LaFeAsO ($\sim$0.25) \cite{Nakamura}, and a sodium-cluster loaded sodalite ($\sim$0.22) \cite{SOD}.  
On top of that, we observe a remarkable long-range tail decaying as 1/$(\lambda r)$. 
We note that numerical studies for simple Hubbard models at $U/t$$\sim$12 without intersite repulsions are well inside the insulating region \cite{Morita,Kyung} contrary to the metallic behavior of $\kappa$-NCS. 
The results strongly suggest a relevant contribution of the offsite interactions to the low-energy physics of the $\kappa$-type compounds, requiring careful reconsiderations for the effective interactions of the $\kappa$-type compounds.  

To conclude, effective low-energy Hamiltonians of organic compounds are established from first principles with the parameters shown in TABLE~\ref{PARAM} for $\kappa$-NCS and $\kappa$-CN. 
The derived parameters indicate that (i) the geometrical frustration parameter $|t'/t|$ is substantially smaller than the extended H\"uckel results and $\kappa$-CN estimated at $|t'/t|$$\sim$0.8 has turned out to be away from the right triangular structure and (ii) the onsite Coulomb repulsion ($U$$\sim$0.8 eV characterized by $U/t$$\sim$12-15) is unexpectedly large compared to the H\"uckel estimate given by $U/t$$\sim$7-8, while the intersite Coulomb interaction was found to be also appreciable. Reexaminations on the low-energy physics of the $\kappa$-ET compounds are desired on the present quantitative and reliable basis, which will need future studies using accurate solvers for the realistic model.

We thank financial support from MEXT Japan under the grant numbers 16076212, 17071003, 17064004, 19019012, and 19014022. 
All the computations have been performed on Hitachi SR11000 system at the Supercomputer Center, Institute for Solid State Physics, the University of Tokyo and on the same system of Supercomputing Division, Information Technology Center, the University of Tokyo.

\end{document}